\documentclass[aps,prb,twocolumn]{revtex4}
\usepackage{amsmath}
\usepackage{graphicx}
\begin{document}

\title{Spin-Orbit Torques and Anisotropic Magnetization Damping in Skyrmion Crystals}

\author{Kjetil M. D. Hals and Arne Brataas}
\affiliation{ Department of Physics, Norwegian University of Science and Technology, NO-7491, Trondheim, Norway }

\begin{abstract}
The length scale of the magnetization gradients in chiral magnets is determined by the relativistic Dzyaloshinskii-Moriya interaction. Thus, even conventional spin-transfer torques are controlled by the relativistic spin-orbit coupling in these systems, and additional relativistic corrections to the current-induced torques and magnetization damping become important for a complete understanding of the current-driven magnetization dynamics. We theoretically study the effects of reactive and dissipative homogeneous spin-orbit torques and anisotropic damping on the current-driven skyrmion dynamics in cubic chiral magnets. Our results demonstrate that spin-orbit torques play a significant role in the current-induced skyrmion velocity. The dissipative spin-orbit torque generates a relativistic Magnus force on the skyrmions, whereas the reactive spin-orbit torque yields a correction to both the drift velocity along the current direction and the transverse velocity associated with the Magnus force. The spin-orbit torque corrections to the velocity scale linearly with the skyrmion size, which is inversely proportional to the spin-orbit coupling. 
Consequently, the reactive spin-orbit torque correction can be the same order of magnitude as the non-relativistic contribution. More importantly, the dissipative spin-orbit torque can be the dominant force that causes a deflected motion of the skyrmions if the torque exhibits a linear or quadratic relationship with the spin-orbit coupling. In addition, we demonstrate that the skyrmion velocity is determined by anisotropic magnetization damping parameters governed by the skyrmion size.
\end{abstract}
\maketitle

\section{Introduction}
The manipulation of submicron-scale magnetic elements via electric currents has paved the way for a promising new class of magnetoelectronic devices with improved scalability, faster execution time, and lower power consumption.~\cite{Ralph:jmmm08, Brataas:nature2012} The dominant mechanism of current-driven magnetic excitations in conventional metallic ferromagnets is the spin-transfer-torque (STT) effect,~\cite{Berger, Slon} in which spin angular momentum is transferred from spin-polarized currents to the magnetization, generating the torque $\boldsymbol{\tau} (\mathbf{r}, t)$ on the magnetization:~\cite{Ralph:jmmm08, Brataas:nature2012}
\begin{equation}
\boldsymbol{\tau} (\mathbf{r}, t)  =  - \left( 1 - \beta\mathbf{m}\times   \right)  \left( \mathbf{v}_s\cdot\boldsymbol{\nabla} \right) \mathbf{m}. \label{Eq:STT}
\end{equation} 
Here, the vector $\mathbf{v}_s$ is proportional to the out-of-equilibrium current density $\boldsymbol{\mathcal{J}}$ and the spin polarization of the current. The first term in Eq.~\eqref{Eq:STT} describes the reactive STT, whereas the second term, which is proportional to $\beta$, describes the dissipative STT. 

In systems that lack spatial-inversion symmetry, an alternative manner of generating current-induced magnetization torques is via relativistic intrinsic spin-orbit coupling (SOC). Orbital momentum is (via SOC) transferred to the spins, causing a so-called spin-orbit torque (SOT) on the magnetization.~\cite{Bernevig:prb05, Manchon:prb08, Chernyshov:nature09,  Miron:nature10, Garate:prb09, Hals:epl10, Fang:nn2013, Miron:n2011, Miron:nm2011, Pesin:prb2012, Bijl:prb2012, Wang:prl2012, Liu:prl2012, Emori:arxiv2013, Thomas:nnt2013, Garello:nn2013, Kurebayashi:arxiv2013} 
Recently, SOTs have been observed to lead to remarkably efficient current-driven magnetization dynamics in ultra-thin magnetic films and strained ferromagnetic semiconductors.~\cite{Moore:apl2008, Chernyshov:nature09, Garate:prb09, Hals:epl10,Miron:nature10, Fang:nn2013, Miron:n2011, Miron:nm2011, Thiaville:epl2012,Emori:arxiv2013, Thomas:nnt2013,  Garello:nn2013} 
Similar to the STT in Eq.~\eqref{Eq:STT}, such SOTs also have reactive and dissipative contributions.~\cite{Tatara:prb2013, Kurebayashi:arxiv2013} Although the reactive homogeneous SOT is known to scale linearly with the SOC to the lowest order, ~\cite{Manchon:prb08, Chernyshov:nature09} there are only a few theoretical works regarding the dissipative SOT that typically predict it to be smaller.~\cite{Tatara:prb2013} However, a recent experiment and theory demonstrate that the dissipative homogeneous SOT can be of the same order of magnitude as the reactive part.~\cite{Kurebayashi:arxiv2013} SOTs can enable the design of significantly simpler devices because the torques originate from a direct conversion of orbital angular momentum into spin excitations and appear even in homogenous ferromagnets with no external sources of spin-polarized currents. 

Magnetic skyrmions are vortex-like spin configurations that cannot be continuously deformed into a homogeneous magnetic state.~\cite{Fert:nn2013} Experimentally, magnetic skyrmion phases were first reported in bulk chiral magnets,~\cite{Rossler:nature06, Muhlbauer:science09, Tonomura:nanolett12, Seki:science, Kanazawa:science, Jonietz:science10, Yu:natcom12} and more recently, they also have been observed in magnetic thin films.~\cite{Yu:nature10, Heinze:np11, Yu:natmat11, Huang:prl12, Romming:science13, Sampaio:nn13} Under the application of a weak external magnetic field, a chiral magnetic system crystallizes into a two-dimensional lattice of skyrmions in which the magnetic moment at the core of each vortex is antiparallel to the applied field, whereas the peripheral magnetic moments are parallel. In ultra-thin magnetic films, the skyrmion phase becomes more energetically favorable with decreasing film thickness,~\cite{Heinze:np11, Yu:natmat11, Kiselev:jpd11, Huang:prl12, Sampaio:nn13} and for a single atomic layer of Fe, a skyrmion structure at the atomic length scale has been observed even in the absence of an external magnetic field.~\cite{Heinze:np11}  

In the context of spintronic applications, promising characteristics of skyrmions are the extremely low depinning currents~\cite{Jonietz:science10, Yu:natcom12} that are required to move them and the fact that they avoid pinning centers.~\cite{Iwasaki:nc13} A proposed explanation of the latter feature is that the Magnus force acts on the skyrmions, leading to a deflection of their motion.~\cite{Zang:prl11, Iwasaki:nc13} This force is closely linked to the topological Hall effect and can be viewed as the reaction of the fictitious Lorentz force experienced by the itinerant quasiparticles as their spins adiabatically align with the local magnetization direction. 

The underlying physical mechanism that gives rise to the skyrmion phase is the 
Dzyaloshinskii-Moriya interaction (DMI).~\cite{Dzyaloshinsky:jpcs58, Moriya:pr60} The DMI has the same relativistic origin as the homogeneous SOTs observed in ferromagnetic heterostructures and strained ferromagnetic semiconductors.  They both arise from the combined effect of spin-orbit coupling and broken spatial-inversion symmetry. Neglected in earlier works concerning current-induced skyrmion motion, a reactive SOT in chiral magnets was predicted and studied for systems in the helical phase in Ref.~\onlinecite{Hals:prb2013a}. However, the effect of reactive and dissipative SOTs on the skyrmion dynamics in bulk chiral magnets remains unknown. Such studies are important because SOTs can be equally as important as the conventional STTs in Eq.~\eqref{Eq:STT}. This equal importance is because the typical length scale of the spatial variations of the magnetization is determined by the DMI, i.e., the magnetization gradients $\partial_i m_j$ scale as $\partial_i m_j \sim D/J$, where $D$ is the DMI parameter and $J$ is the spin stiffness. Because $D$ is linear in the SOC, even the non-relativistic STT in Eq.~\eqref{Eq:STT} is proportional to the SOC to the lowest order in the relativistic corrections. Thus, the SOTs in chiral magnets can be of the same order of magnitude as the conventional STT. Thus, a complete understanding of the current-driven dynamics of chiral magnets requires a correct treatment of SOC effects on both the current-induced torques and magnetization damping. 

The magnetization dynamics driven by currents or external fields strongly depends on dissipation. In isotropic systems, the damping can typically be assumed to be decoupled from the magnetization direction and its texture. However, this model does not necessarily extend to chiral magnets. First, the broken spatial-inversion symmetry allows for terms that are linear, not only quadratic, in the magnetization gradients. Second, chiral magnets have a preferred direction, and thus, the dissipation is likely not isotropic and independent of the texture structure.

In this paper, we present a theoretical study of the current-induced dynamics of skyrmions that correctly accounts for the effects of SOC on both current-induced torques and magnetization damping. Our results demonstrate that SOC generates reactive and dissipative homogeneous SOTs that lead to important corrections to the drift velocity along the current direction and to the Magnus force. Another essential consequence of the SOC is that the skyrmions experience effective damping and torque parameters that depend on the current direction relative to the crystallographic axes. 

This paper is organized as follows. Section~\ref{Sec:PhenExp} introduces the phenomenology of SOTs that was presented in Ref.~\onlinecite{Hals:prb2013b} and performs  a similar phenomenological expansion for the Gilbert damping tensor. In Section~\ref{Sec:Results}, we apply the phenomenology to the study of current-induced skyrmion dynamics in cubic chiral magnets and derive a collective coordinate description for the skyrmion velocity. Our results are summarized in Section~\ref{Sec:Summary}.   

\section{Phenomenological Expansion} \label{Sec:PhenExp}
We include the SOC effects phenomenologically by deriving an equation for the magnetization dynamics that satisfies the symmetry of the underlying crystal structure. In deducing expressions for the torques and dissipation, we perform a phenomenological expansion of the magnetization-damping tensor and current-induced torques in terms of the magnetization and its gradients. The magnetic system is assumed to satisfy the local approximation, in which the magnetization dynamics depends only on the local properties of the system. In this approximation, the magnetization dynamics can be phenomenologically expressed in terms of the Landau-Lifshitz-Gilbert-Slonczewski (LLGS) equation:
\begin{equation}
\dot{\mathbf{m}} = -\gamma \mathbf{m}  \times \mathbf{H}_{\rm eff} + \mathbf{m}  \times \boldsymbol{\alpha} \dot{ \mathbf{m}}  + \boldsymbol{\tau} . \label{Eq:LLG}
\end{equation}
Here, $\mathbf{m}(\mathbf{r}, t)$ is the unit vector of the magnetization $\mathbf{M}(\mathbf{r}, t)= M_s \mathbf{m}(\mathbf{r}, t)$, $\gamma$ is (minus) the gyromagnetic ratio, and  $\mathbf{H}_{\rm eff}(\mathbf{r}, t)= -\delta F [\mathbf{M}]/ \delta\mathbf{M}$ is the effective field determined by the functional derivative of the magnetic free-energy functional $F [\mathbf{M}]= \int d\mathbf{r} \mathcal{F}$. $ \mathcal{F} (\mathbf{r}, t)$ is the free-energy density. The second term on the right side of Eq.~\eqref{Eq:LLG} describes the magnetization damping, where   
$\boldsymbol{\alpha} (\mathbf{r}, t)$ is a symmetric, positive-definite second-rank tensor that depends on the local magnetization direction and local magnetization gradients. The torque $\boldsymbol{\tau} (\mathbf{r}, t)$
in Eq.~\eqref{Eq:LLG} represents the current-induced torques.

The free-energy density of a magnetic system is, to the second order in the magnetization gradients, 
\begin{eqnarray}
\mathcal{F} &=& J_{ij} \partial_i\mathbf{M}\cdot\partial_j\mathbf{M} + D_{ijk}M_i \partial_j M_k + K^{(1)}_{ij} M_i M_j + \nonumber \\
& & K^{(2)}_{ijkl} M_i M_j M_k M_l - \mathbf{M}\cdot\mathbf{B}.  \label{Eq:FreeEnergy}
\end{eqnarray}
The tensor $J_{ij}$ is the spin stiffness, $D_{ijk}$ describes the DMI, and $\mathbf{B}$ is an external magnetic field. In an inversion-symmetric system, $D_{ijk}= 0$.
The two terms proportional to $K^{(1)}_{ij}$ and $K^{(2)}_{ijkl}$ represent the two first harmonics in the phenomenological expansion of the magnetocrystalline anisotropy energy.      
$J_{ij}$, $D_{ijk}$, $K^{(1)}_{ij}$, and $K^{(2)}_{ijkl}$ are polar tensors that are invariant under the point group of the system.~\cite{comment1} 
In the above equation and in what follows, we assume summation over repeated indices, and $\partial_i$ is a shorthand notation for $\partial / \partial r_i$.

A similar phenomenological expansion can be performed for the Gilbert damping tensor $\boldsymbol{\alpha} (\mathbf{r}, t)$. To the second order in the spatial magnetization gradients, we obtain
\begin{eqnarray}
\alpha_{ij} &=& \alpha^{(0)}_{ij} + \alpha^{(1)}_{ijkl} m_k m_l + \alpha^{(2)}_{ijklp} m_k\partial_l m_p  + \nonumber \\
& & \alpha^{(3)}_{ijklpq} \partial_k m_l\partial_p m_q + \alpha^{(4)}_{ijklpq} m_k \partial_l\partial_p m_q. 
 \label{Eq:GilbertTensor}
\end{eqnarray}
Again, the tensors $\alpha^{(0)}_{ij}$, $\alpha^{(1)}_{ijkl}$, $\alpha^{(2)}_{ijklp}$,  and $\alpha^{(3,4)}_{ijklpq}$ are invariant under the point group of the system, and the tensor $\alpha^{(2)}_{ijklp}$ only appears in systems with broken spatial-inversion symmetry, such as chiral magnets, the focus of this study. Terms that are odd under time reversal do not appear in the expansion because such terms do not represent dissipative processes. 

Recently, in Ref.~\onlinecite{Hals:prb2013b}, we carried out a phenomenological expansion of the current-induced torque $\boldsymbol{\tau} (\mathbf{r}, t)$. The starting point of the derivation is to write down the most general form of the torque in the linear-response regime and in the local approximation:
\begin{equation}
\boldsymbol{\tau} (\mathbf{r}, t) = \mathbf{m}\times \boldsymbol{\eta}\, \boldsymbol{\mathcal{J}}. \label{Eq:SOT_1}
\end{equation}
The second-rank tensor $\left( \boldsymbol{\eta} [\mathbf{m}, \boldsymbol{\nabla} \mathbf{m}]\right)_{ij}$, which we refer to as the \emph{fieldance tensor}, depends on the local magnetization direction and local gradients of the magnetization.
The fieldance tensor contains all information concerning the local torque. The general form of the fieldance tensor is not known in systems with strong SOC, but recent experiments indicate that it has a complicated structure.~\cite{Garello:nn2013, Kurebayashi:arxiv2013, Emori:arxiv2013, Thomas:nnt2013}
Therefore, we expand the fieldance tensor in powers of 
$m_i$ and $\partial_j m_i$ to find simplified expressions for the allowed torques:
\begin{equation}
\eta_{ij} = \Lambda^{(r)}_{ij} + \Lambda^{(d)}_{ijk} m_k + \beta_{ijkl}\partial_k m_l + P_{ijkln}m_k \partial_l m_n  + ...  \label{Eq:eta}
\end{equation}
The first and second terms in Eq.~\eqref{Eq:eta} represent the reactive and dissipative homogeneous SOTs, respectively, that recently have been observed experimentally.~\cite{Chernyshov:nature09,Miron:nature10,Kurebayashi:arxiv2013}
These terms are only present in systems with broken spatial-inversion symmetry. The last two terms in Eq.~\eqref{Eq:eta} represent a generalization of the torque in Eq.~\eqref{Eq:STT}. 
In the non-relativistic limit, in which the system is invariant under separate rotations of the spin space and coordinate space, the  
$\beta_{ijkl}$ term and $P_{ijkln}$ term  are equal to the dissipative and reactive STTs in Eq.~\eqref{Eq:STT}, respectively.~\cite{Hals:prb2013b}

When SOC is included, the forms of the tensors in Eq.~\eqref{Eq:eta} are, as for the free energy and magnetization damping, determined by the crystal symmetry: $ \Lambda^{(r)}_{ij}$ and $P_{ijkln}$  become invariant axial tensors of the point group, whereas 
$\Lambda^{(d)}_{ijk}$ and $\beta_{ijkl}$ become invariant polar tensors of the point group. Higher-order terms in the expansion of the fieldance tensor represent torques with higher degrees of anisotropy. However,  the 
leading-order terms explicitly written in Eq.~\eqref{Eq:eta} provide a sufficient description of the current-driven dynamics for many materials.     

\section{Chiral Magnets and Skyrmion Dynamics} \label{Sec:Results}
We now focus on bulk magnets with cubic B20-type crystal structures. Examples of such magnets include MnSi, FeGe, and (Fe,Co)Si. The symmetry of these systems is described by the tetrahedral point group T (in the Sch\"onflies notation).  

We begin by writing down the specific form of the invariant tensors in systems described by the point group T.~\cite{Birss:book, Fieschi}
The invariant tensors lead to phenomenological expressions for the free energy, Gilbert damping tensor, and current-induced torque given in Sec.~\ref{Sec:PhenExp}.
A collective coordinate description is then applied to model the current-driven dynamics of the magnetic system.   

\subsection{Invariant tensors}
The point group T is the smallest of the five cubic point groups and consists of 12 proper rotation operators:
the identity operator; three two-fold rotation operators about the x, y, and z axes; and eight three-fold rotation operators about the body diagonals of the cube. Because the group only consists of proper rotations, the invariant axial and polar tensors have the same form.~\cite{comment1} In the present study, expressions are required for all tensors written explicitly in Section~\ref{Sec:PhenExp}, with the exception of $\alpha^{(3)}$ and $\alpha^{(4)}$ in Eq.~\eqref{Eq:GilbertTensor} because we expand the Gilbert damping tensor to the lowest order in the magnetization gradients. Thus, we will consider invariant tensors up to the fifth rank.

Invariant second-rank tensors $T_{ij}$ are described by one independent tensor coefficient:
\begin{equation}
T_{ij} = T \delta_{ij} \, . 
\end{equation}
Here, $\delta_{ij}$ is the Kronecker delta. Invariant third-rank tensors are described by two independent tensor coefficients, and their 
non-vanishing elements satisfy the two symmetry relations
\begin{alignat}{2}
T_{xyz} &=\  & T_{yzx} &= T_{zxy},  \\  
T_{xzy} &=\  & T_{yxz} &= T_{zyx}.
\end{alignat}
Invariant fourth-rank tensors are described by seven independent coefficients that satisfy the relations 
\begin{alignat}{2}
T_{xxxx} &=\  & T_{yyyy} &= T_{zzzz}, \label{Eq:RankFourTensors1}  \\  
T_{xxyy} &=\  & T_{yyzz} &= T_{zzxx}\ \ (p=3), \label{Eq:RankFourTensors2} \\
T_{xxzz} &=\  & T_{yyxx} &= T_{zzyy}\ \ (p=3), \label{Eq:RankFourTensors3}
\end{alignat}
where $p=3$ indicates the three different symmetry relations obtained from the expression by holding the first index constant and permuting the 
last three indices. For example, Eq.~\eqref{Eq:RankFourTensors2} also yields the relations $T_{xyyx}  = T_{yzzy} = T_{zxxz}$ and $T_{xyxy}  = T_{yzyz} = T_{zxzx}$.     

Invariant fifth-rank tensors are determined by 20 independent tensor coefficients whose non-vanishing elements satisfy the symmetry relations
\begin{alignat}{2}
T_{xxxyx} &=\  & T_{yyyzx} &= T_{zzzxy}\ \ (p=20). \label{Eq:RankFiveTensors}
\end{alignat}
Here, $p=20$ refers to the 20 independent symmetry relations obtained from Eq.~\eqref{Eq:RankFiveTensors} via permutations of the five indices.  

There are 68 independent tensor coefficients that determine the required free energy, torques, and damping to describe current-induced skyrmion motion. Whereas the number of parameters might appear overwhelming, we demonstrate below that only certain combinations of these parameters appear in the final results, which are more transparent.

 \subsection{Collective coordinate description}\label{SubSec:CollCoord}
 We apply a collective coordinate description to model the current-driven magnetization dynamics.~\cite{Tretiakov:prl2008}
 The magnetic state is assumed to be parameterized by a set of time-dependent collective coordinates $\left\{ a_i (t) | i=1,2,... \right\}$ such that
 $\mathbf{m} (\mathbf{r}, t)=  \mathbf{m} (\mathbf{r}, \left\{ a_i (t) \right\})$.
 The equations of motion for the collective coordinates are then given by
\begin{equation}
 \left( \Gamma_{ij} - G_{ij} \right) \dot{a}_j = \gamma F_i +   L_i . \label{Eq:CC} 
\end{equation} 
Here, the matrices $\Gamma_{ij}$ and $G_{ij}$ are $G_{ij} = \int  \mathbf{m}\cdot [  (\partial \mathbf{m} / \partial a_i)\times (\partial \mathbf{m} / \partial a_j) ] {\rm dV}$ and
$\Gamma_{ij} = \int  (\partial \mathbf{m} / \partial a_i)\cdot \boldsymbol{\alpha} (\partial \mathbf{m} / \partial a_j)  {\rm dV}$, and $F_i$ is the force attributable to the effective field, 
$F_i = \int  \mathbf{H}_{\rm eff}\cdot (\partial \mathbf{m} / \partial a_i)  {\rm dV}$. The force on the collective coordinates attributable to the current-induced torque is represented by  
$L_i =    \int  (\partial \mathbf{m} / \partial a_i) \cdot [  \mathbf{m} \times \boldsymbol{\tau}] {\rm dV} $. We compute the above matrices and vectors, which are governed by the 68 independent tensor coefficients discussed in the previous section, and determine the rate of change of the collective coordinates. 

 \subsection{Equations of motion}
 A current density $\boldsymbol{\mathcal{J}}=  [\mathcal{J}_x, \mathcal{J}_y, \mathcal{J}_z]$ is applied to the system, and
 an external magnetic field is applied along the $z$ axis such that a lattice of skyrmions forms in the $xy$ plane.  
The skyrmion lattice is assumed to be undistorted during the current-driven dynamics. 
In this approximation, we can disregard rotations of the magnetic texture~\cite{Everschor:prb11}, and the skyrmions can be considered a lattice of non-interacting
 particles, where each skyrmion is described by
 \begin{eqnarray}
 m_x &=& \frac{2 q R (y - r_y)}{ (x-r_x)^2 + (y-r_y)^2 + R^2} , \\
 m_y &=& -\frac{2 q R (x - r_x)}{ (x-r_x)^2 + (y-r_y)^2 + R^2} , \\
 m_z &=& -q \frac{(x-r_x)^2 + (y-r_y)^2 - R^2}{ (x-r_x)^2 + (y-r_y)^2 + R^2} . 
 \end{eqnarray}
 The two-dimensional center-of-mass coordinates $r_x$ and $r_y$ are the collective coordinates that describe the dynamical evolution of each skyrmion.  Distortions introduce an additional collective coordinate, which describes the rotational motion of the skyrmions.~\cite{Everschor:prb11} 
 The parameters $R$ and  $q\in \left\{ 1, -1\right\}$ are the size and topological charge of the skyrmion, respectively.

Disregarding terms that are of second order in the Gilbert damping parameters, 
the collective coordinate formulas presented in Section~\ref{SubSec:CollCoord} generate the velocity of the center of mass: 
\begin{eqnarray}
\begin{pmatrix} \label{Eq:EOM1}
\dot{r}_x \\
\dot{r}_y
\end{pmatrix}
&=&
- \begin{pmatrix}
\left( P_y^{\rm eff} + R  \Lambda^{\rm eff}_{r} \right) \mathcal{J}_x   \\
\left( P_x^{\rm eff} + R  \Lambda^{\rm eff}_{r} \right) \mathcal{J}_y
\end{pmatrix} + \\
& & q 
\begin{pmatrix}
-\left( P_y^{\rm eff}\beta_y^{\rm eff} - P_x^{\rm eff}\alpha_y^{\rm eff}   \right) \mathcal{J}_y   \\
\left( P_x^{\rm eff}\beta_x^{\rm eff} - P_y^{\rm eff}\alpha_x^{\rm eff}    \right) \mathcal{J}_x \nonumber
\end{pmatrix} + \\
& & q R
\begin{pmatrix}
-\left(  \Lambda^{\rm eff}_{d,y} -  \Lambda^{\rm eff}_{r}\alpha_y^{\rm eff}  \right) \mathcal{J}_y   \\
\left(  \Lambda^{\rm eff}_{d,x} -  \Lambda^{\rm eff}_{r}\alpha_x^{\rm eff}  \right) \mathcal{J}_x \nonumber
\end{pmatrix} .
\end{eqnarray}
Here, $P_x^{\rm eff}$ and $P_y^{\rm eff}$ ($P_x^{\rm eff}\beta_x^{\rm eff}$ and $P_y^{\rm eff}\beta_y^{\rm eff}$) are effective reactive (dissipative) STT parameters that are linear combinations of the tensor coefficients $P_{ijkln}$ ($\beta_{ijkln}$) 
and $\alpha_x^{\rm eff}$ and $\alpha_y^{\rm eff}$ are effective Gilbert damping parameters that are linear combinations of the tensor coefficients $\alpha^{(0)}_{ij}$, $\alpha^{(1)}_{ijkl}$, and $\alpha^{(2)}_{ijklp}$.
The parameters $\beta_x^{\rm eff}$ and $\beta_y^{\rm eff}$ are dimensionless and determined by the ratio between the dissipative and reactive STT. 
Their magnitude is proportional to the spin-flip rate, which is of second order in the SOC. 
The effective Gilbert damping parameters depend on the skyrmion size $R$ in combination with the $\alpha^{(2)}_{ijklp}$ tensor, i.e., the parameters can be decomposed into two terms,
$\alpha_{x,y}^{\rm eff}= \alpha_{x,y}^{'} + R^{-1}\alpha_{x,y}^{''} $ where $\alpha_{x,y}^{''}$ depends on $\alpha^{(2)}_{ijklp}$.
The effective reactive and dissipative homogeneous SOT parameters $\Lambda^{\rm eff}_{r}$, $\Lambda^{\rm eff}_{d,x}$, and $\Lambda^{\rm eff}_{d,y}$ are proportional to the tensor coefficients of 
$\Lambda^{(r)}_{ij}$ and $\Lambda^{(d)}_{ijk}$.    
The effective parameters in Eq.~\eqref{Eq:EOM1} determine the current-driven magnetization dynamics and can be extracted from experiments.
Their explicit forms and relations to the invariant tensors are given in Appendix \ref{App1}.

To reveal the effects of the SOC, let us compare the expression for the velocity of the center of mass in Eq. (\ref{Eq:EOM1}) with the conventional expression for the velocity in the non-relativistic limit, in which the homogeneous SOTs vanish, and the STT and the Gilbert damping parameters satisfy the symmetry relations
$P^{\rm eff}\equiv P_x^{\rm eff}= P_y^{\rm eff}$,  $\beta^{\rm eff}\equiv \beta_x^{\rm eff}= \beta_y^{\rm eff}$, and $\alpha^{\rm eff}\equiv \alpha_x^{\rm eff}= \alpha_y^{\rm eff}$:
\begin{eqnarray}
\dot{\mathbf{r}}
&=&
- P^{\rm eff}\boldsymbol{\mathcal{J}}  + 
q P^{\rm eff} (\beta^{\rm eff} - \alpha^{\rm eff}) \hat{\mathbf{z}}\times \boldsymbol{\mathcal{J}}. \label{Eq:EOM2}
\end{eqnarray}
Comparing Eqs.~\eqref{Eq:EOM1} and Eq.~\eqref{Eq:EOM2} indicates that the SOC introduces several important effects on the skyrmion dynamics.

First, the equations of motion in Eq.~\eqref{Eq:EOM1} are no longer rotationally symmetric about the $z$ axis. 
Clearly, the effective torque and damping parameters that govern the motion along the $x$ and $y$ axes differ because there are no symmetry operations that relate the two axes.  Thus, different current-driven velocities can be observed for currents applied along the two directions, and a measurement of this velocity anisotropy provides a simple test for investigating the importance of the SOC.
A current along the $z$ axis does not influence the velocity in the linear-response regime.

Second, the SOTs strongly affect the skyrmion motion along the current direction.   
The reactive homogeneous SOT leads to a renormalization of the drift velocity along the current direction that scales linearly with the skyrmion size $R$.
The reason for this linear dependency is that the homogeneous SOTs do not depend on the magnetization gradients. 
Thus, the homogeneous SOTs couple more strongly to larger skyrmions, whose textures are distributed over a larger spatial region. 
Because $\Lambda^{\rm eff}_{r}$ is linear in the SOC, whereas $R$ scales as the inverse of the SOC (and thus the product $R \Lambda^{\rm eff}_{r}$ is independent of the SOC), the reactive SOT contribution
 to the drift velocity can be of the same order of magnitude as the terms that are induced by the reactive STT, i.e., the terms proportional to $P_x^{\rm eff}$ and $P_y^{\rm eff}$. 
 
 Third, SOTs also strongly influence the Magnus force.  The terms proportional to $q$ in Eqs.~\eqref{Eq:EOM1}-\eqref{Eq:EOM2} represent the transverse drift velocity induced by the Magnus force.
 Both the reactive and dissipative homogenous SOTs produce corrections to the Magnus-force-induced motion that scale linearly with $R$.
 Using the same arguments as above,  the reactive SOT yields a transverse drift velocity $\sim R \Lambda^{\rm eff}_{r}\alpha_{i}^{\rm eff}$ that can be of the same order of magnitude as the 
 non-relativistic terms $\sim P_{i}^{\rm eff}\beta_{i}^{\rm eff}$ and $\sim P_{i}^{\rm eff}\alpha_{i}^{\rm eff}$ ($i\in \left\{ x, y\right\}$) if $\beta_{i}^{\rm eff} << 1$.
 The most interesting observation from Eq.~\eqref{Eq:EOM1} is the contribution of the dissipative homogeneous SOT to the transverse velocity. 
 In contrast to the terms that arise from the STTs and reactive SOT, the dissipative SOT produces a transverse velocity that depends neither on damping parameters nor the dimensionless effective $\beta$ parameters.
 The velocity is solely determined by the values of $R\Lambda^{\rm eff}_{d,x}$ and $R\Lambda^{\rm eff}_{d,y}$.    
 There is little knowledge regarding the magnitude of the dissipative homogeneous SOT and how it depends on the SOC in chiral magnets. However, a recent experiment concerning (Ga,Mn)As indicated that the dissipative part can
 be comparable in magnitude to the reactive part. 
 If  the same result is applicable to chiral magnets, the dissipative homogeneous SOT provides the largest contribution to the transverse drift velocity and is the \emph{ dominant driving force} that causes deflected motion of the skyrmions.
 This relativistic Magnus force is not linked to the fictitious magnetic field generated by the spin texture but instead arises from the dissipative part of the out-of-equilibrium spin density generated by the SOC combined with an applied electric field.

The three independent tensor coefficients that describe the reactive and dissipative homogeneous SOTs can be extracted from spin-orbit ferromagnetic resonance (FMR) measurements.~\cite{Fang:nn2013,Kurebayashi:arxiv2013}
An external magnetic field is used to align the magnetization along different directions relative to the bar direction, and an alternating current is applied to produce microwave SOTs within the sample that resonantly drive the magnetization. 
The reflected direct current contains information regarding the magnitude of the SOTs.    
We believe that such a measurement of the SOTs will be one of most interesting tasks for future experimental work concerning chiral magnets.
 
Just prior to the submission of our paper, a related theoretical work concerning SOTs and skyrmions in magnetic thin films was presented.~\cite{Knoester:arXiv13} However, that work considers a different symmetry class that is intended for the description of ultra-thin ferromagnetic heterostructures, in which there is complete rotational symmetry and broken spatial-inversion symmetry along a transverse direction. 
Both reactive and dissipative Rashba SOTs are considered in their study, and they demonstrate that the SOTs also play a significant role in the skyrmion dynamics of these systems. However, only isotropic and spatially independent damping is considered.    
 
\section{Summary} \label{Sec:Summary}
In summary, we studied the effects of SOC on the current-driven dynamics of skyrmions in cubic chiral magnets. 
We performed a phenomenological expansion of the Gilbert damping tensor and current-induced torques that accounts for the relativistic SOC effects.
A collective-coordinate description was applied to model the current-induced motion of the skyrmions.
Our results demonstrated that the skyrmion velocity depends on the direction of the applied current relative to the crystallographic axes and that the SOTs 
contribute significantly to the current-induced velocity. The reactive SOT induces a correction to both the parallel and transverse drift velocities of the skyrmions that is of the same order of magnitude
as the non-relativistic contributions. If the dissipative SOT exhibits a linear or quadratic relationship with the SOC, it produces a relativistic Magnus-force motion that is larger than the transverse drift velocity induced by conventional STTs.  
The SOTs cannot be neglected in the modeling of current-driven skyrmion dynamics because they do not depend on the gradients of the magnetization and couple more strongly to larger skyrmions.

\section{Acknowledgments}
K.M.D.H. would like to thank Rembert Duine for stimulating discussions of SOTs in chiral magnets.
K.M.D.H. and A.B. would like to thank Jairo Sinova for notifying us of Ref.~\onlinecite{Bernevig:prb05} and its 
important contribution to the theory of SOTs.

\appendix

\section{Effective Parameters}\label{App1}
Below, we provide the expressions for the effective torque and damping parameters in terms of the tensor coefficients of the invariant tensors given in Section~\ref{Sec:PhenExp}. The effective damping parameters are
\begin{widetext}
\begin{eqnarray}
\alpha_x^{\rm eff} &=& \frac{1}{60} (60 a^{(0)} + 12 a^{(1)}_{xxxx} + 19 a^{(1)}_{xxyy} + 29 a^{(1)}_{xxzz} - 10 a^{(1)}_{xyxy} - 10 a^{(1)}_{xyyx} - 2 a^{(1)}_{xzxz} - 2 a^{(1)}_{xzzx} )  - \nonumber \\
& & \frac{1}{30 R} (3 a^{(2)}_{xxxyz} - 5 a^{(2)}_{xxxzy} - 9 a^{(2)}_{xxyxz} + 10 a^{(2)}_{xxyzx} + 7 a^{(2)}_{xxzxy} -     6 a^{(2)}_{xxzyx} + 8 a^{(2)}_{xyxxz} - 6 a^{(2)}_{xyxzx} -   \nonumber \\ 
& &   2 a^{(2)}_{xyyyz} +   4 a^{(2)}_{xyyzy}  -  2 a^{(2)}_{xyzxx} + 2 a^{(2)}_{xyzzz} + 2 a^{(2)}_{xzxxy} - 2 a^{(2)}_{xzxyx} - 2 a^{(2)}_{xzyxx} + 2 a^{(2)}_{xzyyy}  ),   \\
\alpha_{y}^{\rm eff}  &=&   \frac{1}{60} (60 a^{(0)} + 12 a^{(1)}_{xxxx} + 29 a^{(1)}_{xxyy} + 19 a^{(1)}_{xxzz} -  6 a^{(1)}_{xyxy} - 6 a^{(1)}_{xyyx} - 6 a^{(1)}_{xzxz} - 6 a^{(1)}_{xzzx} ) -  \nonumber \\ 
         & & \frac{1}{30 R} (5 a^{(2)}_{xxxyz} - 3 a^{(2)}_{xxxzy} - 7 a^{(2)}_{xxyxz} + 6 a^{(2)}_{xxyzx} +  9 a^{(2)}_{xxzxy} -  10 a^{(2)}_{xxzyx} + 2 a^{(2)}_{xyxzx} - 2 a^{(2)}_{xyyyz} +  \nonumber \\
         & &    2 a^{(2)}_{xyzxx} - 2 a^{(2)}_{xyzzz} - 6 a^{(2)}_{xzxxy} + 6 a^{(2)}_{xzxyx} + 2 a^{(2)}_{xzyxx} - 2 a^{(2)}_{xzyyy} -  4 a^{(2)}_{xzzyz}  )  .        
\end{eqnarray}
The effective polarizations are
\begin{eqnarray}
P_x^{\rm eff} &=&  -\frac{1}{12}(P_{xxxzy} + 3 P_{xxyxz} - P_{xxyzx} - 3 P_{xxzxy} - P_{xyxxz} + P_{xyyyz} + P_{xyzxx} -  P_{xyzyy} - P_{xzyyy} + \nonumber \\
& &  P_{xzyzz} +  P_{xzzyz} - 3 P_{xzzzy}),  \\
P_y^{\rm eff} &=& \frac{1}{12}(P_{xxxyz} - 3 P_{xxyxz} + 3 P_{xxzxy} - P_{xxzyx} - 3 P_{xyyyz} + P_{xyyzy} + P_{xyzyy} -  P_{xyzzz} - P_{xzxxy} + \nonumber \\
& &  P_{xzyxx} - P_{xzyzz} + P_{xzzzy}).
\end{eqnarray}
The effective $\beta$ factors are defined via the expressions
\begin{eqnarray}
P_x^{\rm eff}\beta_x^{\rm eff} &=& -\frac{1}{6}(\beta_{xxxx} - \beta_{xxyy} + 2 \beta_{xyyx} - \beta_{xzxz} + 3 \beta_{xzzx}), \\
P_y^{\rm eff}\beta_y^{\rm eff} &=& \frac{1}{6}(-\beta_{xxxx} + \beta_{xxzz} + \beta_{xyxy} - 3 \beta_{xyyx} - 2 \beta_{xzzx}),   
\end{eqnarray}   
and finally, the effective SOT parameters are given by
\begin{eqnarray}
\Lambda^{\rm eff}_{r}&=& -\Lambda^{(r)}_{xx} / 2, \  \  \Lambda^{\rm eff}_{d,x}= -\Lambda^{(d)}_{xyz} / 2,\  \  \Lambda^{\rm eff}_{d,y}= \Lambda^{(d)}_{xzy} / 2.             
\end{eqnarray}     
These expressions are used  to quantify the center-of-mass velocity in Eq.~ \eqref{Eq:EOM1}.
\end{widetext}


\begin{thebibliography}{99}
\bibitem{comment1} A polar tensor $T_{ij\ldots k}$ is a tensor that transforms as 
$T_{ij\ldots k}= R_{i\alpha} R_{j\beta}\ldots R_{k\gamma}T_{\alpha\beta\ldots \gamma}$ under a coordinate transformation $\mathbf{x'}=\mathbf{Rx}$. 
An axial tensor is a tensor that transforms as 
$T_{ij\ldots k}= |R| R_{i\alpha} R_{j\beta}\ldots R_{k\gamma}T_{\alpha\beta\ldots \gamma}$,
where $|R|$ is the determinant of $\mathbf{R}$.
\end{thebibliography}
\end{document}